\documentclass[twocolumn,nofootinbib]{revtex4-2}
\usepackage{amsmath,amssymb,bm,graphicx,hyperref}
\allowdisplaybreaks[1]

\newcommand\+{\dagger}

\renewcommand\>{\rangle}
\newcommand\up{\uparrow}
\newcommand\down{\downarrow}
\newcommand\eps{\varepsilon}
\newcommand\0{{\bm{0}}}
\renewcommand\k{{\bm{k}}}
\newcommand\p{{\bm{p}}}
\newcommand\q{{\bm{q}}}
\newcommand\s{{\bm{s}}}
\renewcommand\S{{\bm{S}}}
\newcommand\ex{\mathrm{ex}}
\newcommand\FS{\mathrm{FS}}
\DeclareMathOperator\Tr{Tr}

\begin{document}

\title{Kondo effect under arbitrary spin-momentum locking}

\author{Kinari Goto}
\author{Yusuke Nishida}
\affiliation{Department of Physics, Institute of Science Tokyo,
Ookayama, Meguro, Tokyo 152-8551, Japan}

\date{June 2025}

\begin{abstract}
The Kondo effect originates from the spin exchange scattering of itinerant electrons with a localized magnetic impurity.
Here, we consider generalization of Weyl-type electrons with their spin locked on a spherical Fermi surface in an arbitrary way and study how such spin-momentum locking affects the Kondo effect.
After introducing a suitable model Hamiltonian, a simple formula for the Kondo temperature is derived with the second-order perturbation theory, which proves to depend only on the spin averaged over the Fermi surface.
In particular, the Kondo temperature is unaffected as long as the average spin vanishes, but decreases as the average spin increases in its magnitude, and eventually vanishes when the spin is completely polarized on the Fermi surface, illuminating the role of spin-momentum locking in the Kondo effect.
\end{abstract}

\maketitle

\section{Introduction}
The Kondo effect is one of the most fundamental quantum phenomena in many-body physics, where the interaction of fermions with a localized impurity is enhanced logarithmically toward low temperature, eventually leading to the formation of a screening cloud below the Kondo temperature~\cite{Hewson:1993,Coleman:2007}.
Whereas the Kondo effect was originally discovered in explaining an anomalous electrical resistivity minimum in dilute magnetic alloys~\cite{Kondo:1964}, it has been observed also in artificial nanosystems such as quantum dots~\cite{Goldhaber-Gordon:1998,Cronenwett:1998,Schmid:1998}, carbon nanotubes~\cite{Nygard:2000,Buitelaar:2002}, and individual molecules~\cite{Park:2002,Liang:2002}.
Furthermore, realization of the Kondo effect has been proposed theoretically in ultracold atomic gases~\cite{Nishida:2013,Bauer:2013,Nakagawa:2015} (see references in Ref.~\cite{Nishida:2016} for earlier proposals) and even in quark and nuclear matter~\cite{Yasui:2013,Hattori:2015,Yasui:2016}.

Microscopic origin of the Kondo effect lies in the coherent spin exchange between a localized magnetic impurity and a surrounding Fermi sea of itinerant electrons.
Here, the spin degeneracy on the Fermi surface is essential and the Kondo temperature reads
\begin{align}\label{eq:kondo}
T_K \propto \exp\!\left(-\frac1{J\nu_F}\right),
\end{align}
with $J$ the antiferromagnetic exchange coupling and $\nu_F$ the total density of states at the Fermi energy~\cite{Hewson:1993,Coleman:2007}.
The role of spin can actually be replaced with another degenerate degree of freedom such as an orbital angular momentum~\cite{Cox:1998}, leading to the orbital Kondo
effect~\cite{Kolesnychenko:2002,Jarillo-Herrero:2005}.
A natural question then may be what if the spin and orbital degeneracies are lost by coupling spin and orbital degrees of freedom.
Dirac and Weyl semimetals provide such examples and considerable interest has been devoted to the Kondo effect therein~\cite{Principi:2015,Yanagisawa:2015,Mitchell:2015,Sun:2015,Lu:2019,Pedrosa:2021} as well as under Rashba spin-orbit interactions~\cite{Malecki:2007,Zitko:2011,Zarea:2012,Mastrogiuseppe:2014,Wong:2016,Sousa:2016}.

The purpose of our work is to consider generalization of Weyl-type electrons with their spin locked on a spherical Fermi surface in an arbitrary way and study how such spin-momentum locking affects the Kondo effect.
A naive expectation may be that the Kondo effect is to be disfavored because the spin exchange scattering is restricted in the momentum space.
However, somewhat unexpectedly, we will show that the Kondo temperature is unaffected as long as the spin averaged over the Fermi surface vanishes.
Toward this end, after introducing our model Hamiltonian in Sec.~\ref{sec:model}, the scattering problem is analyzed in Sec.~\ref{sec:scattering} with the second-order perturbation theory.
We set $\hbar=k_B=1$ throughout this paper and employ shorthand notation $\int_\p\equiv\int d^d\p/(2\pi)^d$ for $d$-dimensional momentum integration.

\section{Model Hamiltonian}\label{sec:model}
Weyl-type electrons with arbitrary spin-momentum locking are described by
\begin{align}\label{eq:hamiltonian}
H_0 = \sum_{\alpha,\beta}\int_\p\,c^\+_{\alpha\p}
\left[vp\s(\hat\p)\cdot\bm\sigma_{\alpha\beta}
- \mu\delta_{\alpha\beta}\right]c_{\beta\p},
\end{align}
where $v$ is the Fermi velocity, $\mu$ is the chemical potential, $\bm\sigma=(\sigma_x,\sigma_y,\sigma_z)$ are the Pauli matrices, and $\s(\hat\p)$ is an arbitrary unit vector depending on the direction of momentum $\p$.
A single electron state has the positive or negative energy $\eps=\pm vp-\mu$ when its spin is parallel to $\pm\s(\hat\p)$.
Therefore, the direction of spin is locked to $\pm\s(\hat\p)$ for a given momentum, which obviously generalizes the case of $\s(\hat\p)=\pm\hat\p$ for Weyl semimetals~\cite{Principi:2015,Yanagisawa:2015,Mitchell:2015,Sun:2015}.
Since the dispersion relation is assumed to be isotropic and linear, our Hamiltonian may not directly apply to multi-Weyl semimetals with anisotropic and nonlinear dispersion relations~\cite{Lu:2019,Pedrosa:2021}, but is suitable to focus on the role of spin-momentum locking.
For further generality, the spatial dimension is also allowed to be arbitrary, including $d=1$, 2, and 3.

In order to study the Kondo effect, we couple itinerant electrons with a magnetic impurity localized at the origin via
\begin{align}\label{eq:exchange}
H_\ex = J\sum_{\alpha,\beta}\int_{\p,\q}
\S\cdot c^\+_{\alpha\p}\bm\sigma_{\alpha\beta}c_{\beta\q}.
\end{align}
This is the spin exchange interaction with $\S=(S_x,S_y,S_z)$ assumed to be spin-1/2 operators, which can be derived in the low-energy limit of the Anderson impurity model as shown in Appendix~\ref{sec:derivation}.
The total Hamiltonian then reads $H=H_0+H_\ex$.

For our analysis, it is useful to introduce the basis that diagonalizes the single electron Hamiltonian in Eq.~(\ref{eq:hamiltonian}) as
\begin{align}\label{eq:unitary}
\s(\hat\p)\cdot\bm\sigma = U_\s(\hat\p)^\+\sigma_3U_\s(\hat\p),
\end{align}
where $U_\s(\hat\p)$ is the $2\times2$ unitary matrix provided by
\begin{align}\label{eq:matrix}
U_\s(\hat\p) = \frac{s_x(\hat\p)\sigma_x + s_y(\hat\p)\sigma_y
+ [1+s_z(\hat\p)]\sigma_z}{\sqrt{2+2s_z(\hat\p)}}.
\end{align}
The creation and annihilation operators in the new basis are
\begin{align}
c_{a\p}^\+ = \sum_\alpha c_{\alpha\p}^\+[U_\s(\hat\p)^\+]_{\alpha a}, \quad
c_{b\p} = \sum_\beta U_\s(\hat\p)_{b\beta}c_{\beta\p},
\end{align}
respectively, with $\alpha$ and $\beta$ valued on $\up,\down$ and $a$ and $b$ on $+,-$.
Consequently, the Hamiltonian for itinerant electrons is turned into
\begin{align}
H_0 = \sum_a\int_\p\,\eps_{ap}c^\+_{a\p}c_{a\p},
\end{align}
whereas their spin exchange interaction with the localized magnetic impurity into
\begin{align}
H_\ex = J\sum_{a,b}\int_{\p,\q}
\S\cdot c^\+_{a\p}\bm\Sigma_\s(\hat\p,\hat\q)_{ab}c_{b\q},
\end{align}
where $\eps_{ap}\equiv avp-\mu$ is the single electron energy and
\begin{align}\label{eq:sigma}
\bm\Sigma_\s(\hat\p,\hat\q) \equiv U_\s(\hat\p)\bm\sigma U_\s(\hat\q)^\+.
\end{align}
It is important to note that the spin-momentum locking is now encoded in $\bm\Sigma_\s(\hat\p,\hat\q)$, which do not obey the commutation relations of Pauli matrices unless $\s(\hat\p)=\s(\hat\q)$.
We also introduce $\Sigma_\s^0(\hat\p,\hat\q)\equiv U_\s(\hat\p)U_\s(\hat\q)^\+$ for later use.

\section{Scattering problem}\label{sec:scattering}
\subsection{$T$ matrix}
The $T$ matrix of itinerant electrons scattered by the localized magnetic impurity is provided by
\begin{align}
T(E) = H_\ex + H_\ex\frac1{E-H_0+i0^+}H_\ex + O(J^3)
\end{align}
up to the second order in the exchange coupling.
We are interested in its matrix element $\<a\p|T(E)|a\q\>$, where $|a\p\>=c_{a\p}^\+|\FS\>$ represents an electron added on top of the Fermi sea, with $a=+\,(-)$ for $\mu\geq0\,(<0)$.
Its energy reads $E=E_\FS+\eps_{ap}$, where $E_\FS$ is that of the Fermi sea and $\eps\equiv\eps_{ap}=\eps_{aq}>0$ due to the energy conservation.
The first-order term is simply
\begin{align}
\<a\p|H_\ex|a\q\> = J\,\S\cdot\bm\Sigma_\s(\hat\p,\hat\q)_{aa}.
\end{align}

\begin{figure}[t]
\includegraphics[width=\columnwidth,clip]{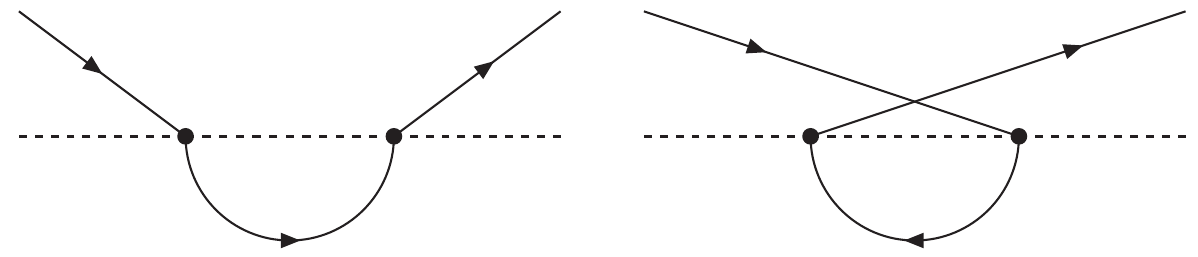}
\caption{\label{fig:perturbation}
Feynman diagrams for the $T$ matrix to the second order in perturbation.
Solid curves, dashed lines, and dots represent itinerant electrons, localized magnetic impurity, and spin exchange interaction, respectively.}
\end{figure}

The second-order term has two contributions depicted by the Feynman diagrams in Fig.~\ref{fig:perturbation},
\begin{align}
& \<a\p|H_\ex\frac1{E-H_0+i0^+}H_\ex|a\q\> \notag\\
&= J^2\sum_b\int_\k\,\S\cdot\bm\Sigma_\s(\hat\p,\hat\k)_{ab}
F_{bk}(\eps)\S\cdot\bm\Sigma_\s(\hat\k,\hat\q)_{ba} \notag\\
&\quad + J^2\sum_b\int_\k\,\S\cdot\bm\Sigma_\s(\hat\k,\hat\q)_{ba}
G_{bk}(\eps)\S\cdot\bm\Sigma_\s(\hat\p,\hat\k)_{ab},
\end{align}
where
\begin{align}
F_{bk}(\eps) \equiv \frac{\theta(\eps_{bk})}{\eps-\eps_{bk}+i0^+}, \quad
G_{bk}(\eps) \equiv \frac{\theta(-\eps_{bk})}{\eps-\eps_{bk}-i0^+}
\end{align}
are the propagators of electron and hole, respectively.
We then employ $F_{bk}(\eps)=[\{F_{+k}(\eps)+F_{-k}(\eps)\}\sigma_0/2+\{F_{+k}(\eps)-F_{-k}(\eps)\}\sigma_3/2]_{bb}$, the same for $G_{bk}(\eps)$, Eqs.~(\ref{eq:sigma}), and (\ref{eq:unitary}) to obtain
\begin{align}
& \<a\p|H_\ex\frac1{E-H_0+i0^+}H_\ex|a\q\> \notag\\
&= J^2\int_\k\,\sum_b\frac{F_{bk}(\eps)}{2}S_iS_j
[U_\s(\hat\p)\sigma_i\sigma_jU_\s(\hat\q)^\+]_{aa} \notag\\
&\quad + J^2\int_\k\,\sum_bb\frac{F_{bk}(\eps)}{2}S_iS_j\,
\s(\hat\k)\cdot[U_\s(\hat\p)\sigma_i\bm\sigma\sigma_jU_\s(\hat\q)^\+]_{aa} \notag\\
&\quad + J^2\int_\k\,\sum_b\frac{G_{bk}(\eps)}{2}S_jS_i
[U_\s(\hat\p)\sigma_i\sigma_jU_\s(\hat\q)^\+]_{aa} \notag\\
&\quad + J^2\int_\k\,\sum_bb\frac{G_{bk}(\eps)}{2}S_jS_i\,
\s(\hat\k)\cdot[U_\s(\hat\p)\sigma_i\bm\sigma\sigma_jU_\s(\hat\q)^\+]_{aa},
\end{align}
with the summations over $i$ and $j$ implictly assumed.
Finally, $\S$ being spin-1/2 and the separation of radial and angular integrations of $\k$ lead to
\begin{align}
& \<a\p|H_\ex\frac1{E-H_0+i0^+}H_\ex|a\q\> \notag\\
&= \frac{3J^2}{4}I_1(\eps)\,\Sigma_\s^0(\hat\p,\hat\q)_{aa}
- \frac{J^2}{4}I_2(\eps)\,\bar\s\cdot\bm\Sigma_\s(\hat\p,\hat\q)_{aa} \notag\\
&\quad - J^2I_3(\eps)\,\S\cdot\bm\Sigma_\s(\hat\p,\hat\q)_{aa}
+ J^2I_4(\eps)\,\S\cdot\bar\s\ \Sigma_\s^0(\hat\p,\hat\q)_{aa}.
\end{align}
Here, the spin averaged over the Fermi surface
\begin{align}
\bar\s \equiv \frac1{A_{d-1}}\int d\Omega_{\hat\k}\,\s(\hat\k)
\end{align}
is introduced with $A_{d-1}=\int d\Omega_{\hat\k}=2\pi^{d/2}/\Gamma(d/2)$ as well as four kinds of integrals
\begin{align}
I_1(\eps) &\equiv \frac{A_{d-1}}{(2\pi)^d}\int_0^\Lambda\!dk\,k^{d-1}
\sum_b\frac{F_{bk}(\eps)+G_{bk}(\eps)}{2}, \\
I_2(\eps) &\equiv \frac{A_{d-1}}{(2\pi)^d}\int_0^\Lambda\!dk\,k^{d-1}
\sum_bb\frac{F_{bk}(\eps)+G_{bk}(\eps)}{2}, \\
I_3(\eps) &\equiv \frac{A_{d-1}}{(2\pi)^d}\int_0^\Lambda\!dk\,k^{d-1}
\sum_b\frac{F_{bk}(\eps)-G_{bk}(\eps)}{2}, \\
I_4(\eps) &\equiv \frac{A_{d-1}}{(2\pi)^d}\int_0^\Lambda\!dk\,k^{d-1}
\sum_bb\frac{F_{bk}(\eps)-G_{bk}(\eps)}{2}
\end{align}
with the momentum cutoff $\Lambda$.
Whereas $I_1(\eps)$ and $I_2(\eps)$ are finite in the limit of $\eps\to0$, $I_3(\eps)$ and $I_4(\eps)$ diverge logarithmically as
\begin{align}
\lim_{\eps\to0}I_3(\eps) &= \nu_F\ln\eps + (\mathrm{finite}), \\
\lim_{\eps\to0}I_4(\eps) &= a\nu_F\ln\eps + (\mathrm{finite}),
\end{align}
where $\nu_F=A_{d-1}|\mu|^{d-1}/(2\pi v)^d$ is the density of states at the Fermi energy.
We also note that the average spin is restricted to $0\leq|\bar\s|\leq1$ in its magnitude.

Consequently, the $T$ matrix up to the second order in the exchange coupling is found to be
\begin{align}\label{eq:t-matrix}
& \<a\p|T(E)|a\q\> = J\,\S\cdot\bm\Sigma_\s(\hat\p,\hat\q)_{aa} \notag\\
&\quad + \frac{3J^2}{4}I_1(\eps)\,\Sigma_\s^0(\hat\p,\hat\q)_{aa}
- \frac{J^2}{4}I_2(\eps)\,\bar\s\cdot\bm\Sigma_\s(\hat\p,\hat\q)_{aa} \notag\\
&\quad - J^2I_3(\eps)\,\S\cdot\bm\Sigma_\s(\hat\p,\hat\q)_{aa}
+ J^2I_4(\eps)\,\S\cdot\bar\s\ \Sigma_\s^0(\hat\p,\hat\q)_{aa} \notag\\
&\quad + O(J^3).
\end{align}
The resulting $T$ matrix is already useful to study the Kondo effect in the two limiting cases.
When the average spin vanishes, it is reduced to
\begin{align}
& \<a\p|T(E)|a\q\>|_{\bar\s=\0}
= \frac{3J^2}{4}I_1(\eps)\,\Sigma_\s^0(\hat\p,\hat\q)_{aa} \notag\\
&\quad + J[1 - JI_3(\eps)]\,\S\cdot\bm\Sigma_\s(\hat\p,\hat\q)_{aa} + O(J^3).
\end{align}
Here, the second term on the right-hand side indicates that the exchange coupling is renormalized under the poor man's scaling as
\begin{align}
\frac{d J}{d\ln\eps} &= -J^2\nu_F + O(J^3) \\
\Rightarrow\quad
J(\eps) &= \frac{J}{1+J\nu_F\ln\eps+\cdots},
\end{align}
which is enhanced logarithmically toward low energy~\cite{Hewson:1993,Altland-Simons}.
The characteristic scale at which the perturbation theory breaks down defines the Kondo temperature as
\begin{align}\label{eq:s=0}
T_K|_{\bar\s=\0} \propto \exp\!\left(-\frac1{J\nu_F}\right).
\end{align}
This formula coincides with Eq.~(\ref{eq:kondo}) and proves to be unaffected by the spin-momentum locking as long as the average spin vanishes.

On the other hand, when the spin is completely polarized to a particular direction such as $\s(\hat\p)=\s_0$, the $T$ matrix is reduced to
\begin{align}
& \<a\p|T(E)|a\q\>|_{\bar\s=\s_0}
= \frac{J^2}{4}[3I_1(\eps) - aI_2(\eps)] \notag\\
&\quad + aJ[1 - J\{I_3(\eps) - aI_4(\eps)\}]\,\S\cdot\s_0 + O(J^3)
\end{align}
because $\bar\s=\s_0$, $\Sigma_\s^0(\hat\p,\hat\q)=1$, and $\bm\Sigma_{\s_0}=\s_0\sigma_3$ follow from Eqs.~(\ref{eq:unitary}) and (\ref{eq:sigma}).
Whereas $I_3(\eps)$ and $I_4(\eps)$ diverge individually, $I_3(\eps)-aI_4(\eps)$ is finite in the limit of $\eps\to0$.
Therefore, the logarithmic enhancement toward low energy cancels out, indicating
\begin{align}\label{eq:s=1}
T_K|_{\bar\s=\s_0} = 0.
\end{align}
Such an absence of the Kondo effect is actually expected because the spin exchange scattering by the localized magnetic impurity is impossible if the spin of itinerant electrons is completely polarized on the Fermi surface.

\subsection{Scattering rate}
\begin{figure}[t]
\includegraphics[width=0.85\columnwidth,clip]{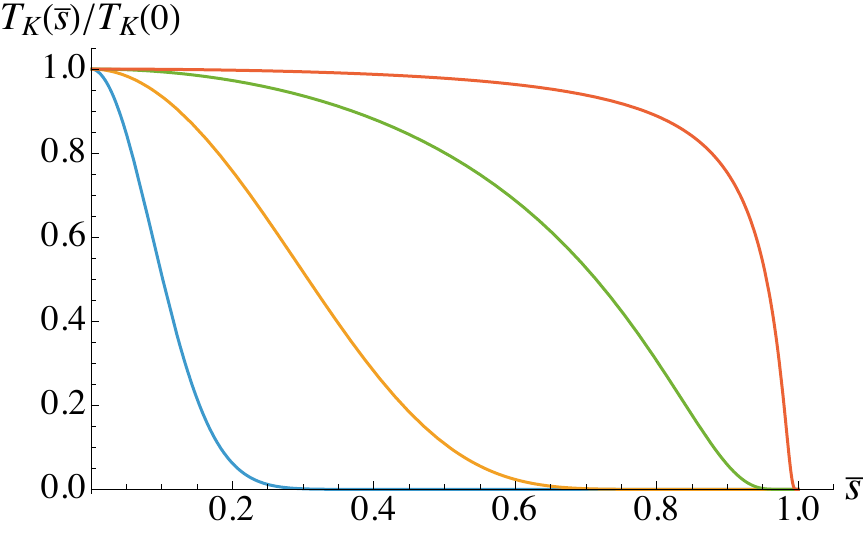}
\caption{\label{fig:temperature}
Normalized Kondo temperature $T_K(\bar\s)/T_K(\0)$ from Eq.~(\ref{eq:temperature}) as a function of $|\bar\s|$ for $J\nu_F=1/100$, $1/10$, $1$, and $10$ from the leftmost curve to the rightmost one.}
\end{figure}

In order to study the Kondo effect beyond the two limiting cases, we analyze
\begin{align}
\gamma(\eps) = \frac1{2A_{d-1}^2}
\iint d\Omega_{\hat\p}\,d\Omega_{\hat\q}\Tr[|\<a\p|T(E)|a\q\>|^2],
\end{align}
which multiplied by the density of final states provides the scattering rate averaged over the initial states.
With the $T$ matrix in Eq.~(\ref{eq:t-matrix}) employed, the trace over the impurity spin is evaluated as
\begin{align}
& \frac12\Tr[|\<a\p|T(E)|a\q\>|^2] \notag\\
&= \frac{J^2}{4}\bm\Sigma_\s(\hat\q,\hat\p)_{aa}
\cdot\bm\Sigma_\s(\hat\p,\hat\q)_{aa} \notag\\
&\quad - \frac{J^3}{2}I_3(\eps)\,\bm\Sigma_\s(\hat\q,\hat\p)_{aa}
\cdot\bm\Sigma_\s(\hat\p,\hat\q)_{aa} \notag\\
&\quad + \frac{J^3}{4}I_4(\eps)\,\bm\Sigma_\s(\hat\q,\hat\p)_{aa}
\cdot\bar\s\ \Sigma_\s^0(\hat\p,\hat\q)_{aa} \notag\\
&\quad + \frac{J^3}{4}I_4(\eps)\,\Sigma_\s^0(\hat\q,\hat\p)_{aa}\,
\bar\s\cdot\bm\Sigma_\s(\hat\p,\hat\q)_{aa} + O(J^4).
\end{align}
Because
\begin{align}
& \bm\Sigma_\s(\hat\q,\hat\p)_{aa}\cdot\bm\Sigma_\s(\hat\p,\hat\q)_{aa}
= \frac{3-\s(\hat\p)\cdot\s(\hat\q)}{2}, \\
& \bm\Sigma_\s(\hat\q,\hat\p)_{aa}\cdot\bar\s\ \Sigma_\s^0(\hat\p,\hat\q)_{aa}
+ \Sigma_\s^0(\hat\q,\hat\p)_{aa}\,\bar\s\cdot\bm\Sigma_\s(\hat\p,\hat\q)_{aa} \notag\\
&= a\bar\s\cdot[\s(\hat\p)+\s(\hat\q)]
\end{align}
follow from Eqs.~(\ref{eq:matrix}) and (\ref{eq:sigma}), the angular integrations of $\p$ and $\q$ then lead to
\begin{align}
& \gamma(\eps) \notag\\
&= \frac{3-\bar\s^2}{8}J^2\left[1 - J\left\{I_3(\eps)
- \frac{2\bar\s^2}{3-\bar\s^2}aI_4(\eps)\right\}\right]^2 + O(J^4) \notag\\
&\to \frac{3-\bar\s^2}{8}J^2\left[1 - \frac{3-3\bar\s^2}{3-\bar\s^2}J\nu_F\ln\eps
+ (\mathrm{finite})\right]^2 + O(J^4)
\end{align}
in the limit of $\eps\to0$.
The resulting scattering rate is enhanced logarithmically toward low energy and the characteristic scale at which the perturbation theory breaks down now reads
\begin{align}\label{eq:temperature}
T_K(\bar\s) \propto \exp\!\left(-\frac{3-\bar\s^2}{3-3\bar\s^2}\frac1{J\nu_F}\right).
\end{align}
Therefore, we find that the Kondo temperature depends only on the spin averaged over the Fermi surface, which smoothly interpolates the two limiting cases of $|\bar\s|=0$ in Eq.~(\ref{eq:s=0}) and $|\bar\s|=1$ in Eq.~(\ref{eq:s=1}).
The obtained Kondo temperature normalized by that for $|\bar\s|=0$ is shown in Fig.~\ref{fig:temperature} as a function of $|\bar\s|$ for several choices of $J\nu_F$.

\section{Summary and outlook}
In summary, we studied the Kondo effect for Weyl-type electrons with arbitrary spin-momentum locking with the second-order perturbation theory and showed that the Kondo temperature is provided by the simple formula in Eq.~(\ref{eq:temperature}).
In particular, it proves to depend only on the spin averaged over the Fermi surface regardless of how the directions of spin and momentum are locked.
As long as the average spin vanishes, the Kondo temperature is unaffected and coincides with the textbook formula in Eq.~(\ref{eq:kondo}) under the spin degeneracy.
As the average spin increases in its magnitude, the Kondo temperature decreases because the spin exchange scattering is suppressed, and eventually vanishes when the spin is completely polarized on the Fermi surface.

Our work thus illuminated the role of spin-momentum locking in the Kondo effect and can be extended toward several research directions.
Since the dispersion relation is assumed to be isotropic and linear in our model Hamiltonian, it is desirable to consider more general dispersion relations relevant to multi-Weyl semimetals.
Furthermore, since our analysis is restricted to the weak-coupling physics above the Kondo temperature, it is important to study the strong-coupling physics below the Kondo temperature such as the formation of a screening cloud.
Finally, it is possible to include the interaction between itinerant electrons in one dimension in terms of Tomonaga-Luttinger liquids along the lines of Refs.~\cite{Furusaki:1994,Furusaki:2005}.
Such future work will provide us with further insight into Kondo physics under spin-momentum locking.

\acknowledgments
This work was supported by JSPS KAKENHI Grant No.~JP21K03384.

\appendix
\section{From Anderson to Kondo impurity model}\label{sec:derivation}
The spin exchange interaction in Eq.~(\ref{eq:exchange}) can be derived from the Anderson impurity model even for Weyl-type electrons in Eq.~(\ref{eq:hamiltonian}) with the standard procedure~\cite{Hewson:1993,Altland-Simons}.
This derivation relies on the low-energy limit as shown below, which is satisfied sufficiently close to the Fermi surface and is justified to determine the Kondo temperature in the weak-coupling limit.

The Anderson impurity model is provided by
\begin{align}
H_A &= \sum_{\alpha,\beta}\int_\p c_{\alpha\p}^\+\eps_{\alpha\beta}(\p)c_{\beta\p}
+ V\sum_\alpha\int_\p\left(d_\alpha^\+c_{\alpha\p} + c_{\alpha\p}^\+d_\alpha\right) \notag\\
&\quad + \eps_d\sum_\alpha n_{d\alpha} + Un_{d\up}n_{d\down},
\end{align}
where $\eps_{\alpha\beta}(\p)=vp\s(\hat\p)\cdot\bm\sigma_{\alpha\beta}-\mu\delta_{\alpha\beta}$ in our work and the hybridization $V$ is assumed to be independent of $\p$ for simplicity.
The total wave function can be decomposed as $|\psi\>=\sum_{n=0}^2|\psi_n\>$, where the subscript $n=0,1,2$ represents the occupation number of the impurity site.
In particular, $|\psi_1\>$ obeys the Schr\"odinger equation,
\begin{align}
& \left[H_{11} + H_{10}\frac1{E-H_{00}}H_{01}
+ H_{12}\frac1{E-H_{22}}H_{21}\right]|\psi_1\> \notag\\
&= E|\psi_1\>,
\end{align}
where $H_{nn'}\equiv P_nHP_{n'}$ and $P_n$ is the projection onto the subspace with the occupation number $n$.
Now, in the low-energy limit of $|\eps_{\alpha\beta}(\p)|\ll|\eps_d|,\eps_d+U$, the corrections due to virtual excitations are evaluated as
\begin{align}
H_{10}\frac1{E-H_{00}}H_{01} \simeq \frac{V^2}{\eps_d}
\sum_{\alpha,\beta}\int_{\p,\q}d_\beta^\+c_{\beta\q}c_{\alpha\p}^\+d_\alpha
\end{align}
and
\begin{align}
H_{12}\frac1{E-H_{22}}H_{21} \simeq \frac{V^2}{-\eps_d-U}
\sum_{\alpha,\beta}\int_{\p,\q}c_{\alpha\p}^\+d_\alpha d_\beta^\+c_{\beta\q},
\end{align}
where $E\sim H_{11}+O(V^2)$ and the single electron energy is neglected in the energy denominator.
We then employ the anticommutation relations and $2\delta_{\alpha\delta}\delta_{\beta\gamma}=\bm\sigma_{\alpha\beta}\cdot\bm\sigma_{\gamma\delta}+\delta_{\alpha\beta}\delta_{\gamma\delta}$ to find
\begin{align}
& H_{10}\frac1{E-H_{00}}H_{01}
\simeq -\frac{V^2}{\eps_d}\sum_{\alpha,\beta}\int_{\p,\q}
\S\cdot c_{\alpha\p}^\+\bm\sigma_{\alpha\beta}c_{\beta\q} \notag\\
&\quad - \frac{V^2}{2\eps_d}\sum_{\alpha,\beta}\int_{\p,\q}
n_{d\alpha}c_{\beta\p}^\+c_{\beta\q}
+ \frac{V^2}{\eps_d}\sum_\alpha\int_\p n_{d\alpha}
\end{align}
and
\begin{align}
& H_{12}\frac1{E-H_{22}}H_{21}
\simeq \frac{V^2}{\eps_d+U}\sum_{\alpha,\beta}\int_{\p,\q}
\S\cdot c_{\alpha\p}^\+\bm\sigma_{\alpha\beta}c_{\beta\q} \notag\\
&\quad + \frac{V^2}{2(\eps_d+U)}\sum_{\alpha,\beta}\int_{\p,\q}
(n_{d\alpha}-1)c_{\beta\p}^\+c_{\beta\q}
\end{align}
with $n_{d\alpha}=d_\alpha^\+d_\alpha$ and $\S=\sum_{\alpha,\beta}d_\alpha^\+\bm\sigma_{\alpha\beta}d_\beta/2$.
Consequently, the above Schr\"odinger equation leads to $H_K|\psi_1\>=E|\psi_1\>$, where the resulting Hamiltonian up to irrelevant constants reads
\begin{align}
H_K &= \sum_{\alpha,\beta}\int_\p c_{\alpha\p}^\+\eps_{\alpha\beta}(\p)c_{\beta\p}
+ J\sum_{\alpha,\beta}\int_{\p,\q}
\S\cdot c_{\alpha\p}^\+\bm\sigma_{\alpha\beta}c_{\beta\q} \notag\\
&\quad + J'\sum_{\alpha,\beta}\int_{\p,\q}c_{\alpha\p}^\+c_{\alpha\q}
\end{align}
with
\begin{align}
J = -\frac{V^2}{\eps_d} + \frac{V^2}{\eps_d+U}, \quad
J' = -\frac{V^2}{2\eps_d} - \frac{V^2}{2(\eps_d+U)}.
\end{align}
The first two terms constitute our model Hamiltonian introduced in Sec.~\ref{sec:model}, whereas the last term provides the potential scattering.

\end{document}